\documentclass[prd,showpacs,preprint,superscriptaddress,nofootinbib]{revtex4}

\def\ka{\left.\mid 11 \right\rangle_c}
\def\kb{\left.\mid 88 \right\rangle_c}
\def\kap{\left.\mid 1'1' \right\rangle_c}
\def\kbp{\left.\mid 8'8' \right\rangle_c}
\def\ba{\,_c\left\langle 11 \mid\right.}
\def\bb{\,_c\left\langle 88 \mid\right.}
\def\bap{\,_c\left\langle 1'1' \mid\right.}
\def\bbp{\,_c\left\langle 8'8' \mid\right.}
\def\paap{\,_c\left\langle 11 \mid 1'1' \right\rangle_c}
\def\papa{\,_c\left\langle 1'1'\mid 11 \right\rangle_c}
\def\pabp{\,_c\left\langle 11 \mid 8'8' \right\rangle_c}
\def\papb{\,_c\left\langle 1'1' \mid 88 \right\rangle_c}
\def\pbap{\,_c\left\langle 88 \mid 1'1' \right\rangle_c}
\def\pbpa{\,_c\left\langle 8'8' \mid 11 \right\rangle_c}
\def\pbpb{\,_c\left\langle 8'8' \mid 88 \right\rangle_c}
\def\pbbp{\,_c\left\langle 88 \mid 8'8' \right\rangle_c}

\providecommand{\openone}{\leavevmode\hbox{\small1\kern-3.8pt\normalsize1}}

\usepackage[dvips]{graphicx}
\usepackage{amsmath}
\usepackage{amsfonts}
\usepackage{epsfig}
\usepackage{color}

\begin{document}
\title{Probabilities in nonorthogonal basis: Four--quark systems}

\author{J. Vijande}
\affiliation{Departamento de F\'{\i}sica At\'{o}mica, Molecular y Nuclear, Universidad de Valencia (UV)  
and IFIC (UV-CSIC), Valencia, Spain.}
\author{A. Valcarce}
\affiliation{Departamento de F\'{\i}sica Fundamental, Universidad de Salamanca, Salamanca, Spain.}

\date\today

\begin{abstract}
Four-quark states may exist as colorless meson-meson molecules or compact
systems with two-body colored components.
We derive an analytical procedure to expand an arbitrary four--quark
wave function in terms of nonorthogonal color singlet--singlet vectors. 
Using this expansion we develop the necessary formalism to evaluate the probability 
of physical components with an arbitrary four-quark wave function.
Its application to characterize bound and 
unbound four--quark states as meson-meson, molecular or compact systems is discussed
\end{abstract}
\maketitle

\section{Introduction}

The physics of heavy quarks has become one of the best laboratories exposing the
limitations and challenges of the naive quark model and also giving hints
into a more mature description of hadron spectroscopy. More than thirty years 
after the so-called November revolution~\cite{Bjo85}, heavy meson spectroscopy 
is being again a challenge. Its formerly comfortable world is being severely 
tested by new experiments reporting states that do not fit into a 
simple quark-antiquark configuration~\cite{Ros07}.
It seems nowadays unavoidable to resort to higher order Fock space components to tame the 
bewildering landscape arising with these new findings. Four--quark components, 
either pure or mixed with $q\bar q$ states, constitute a natural explanation for the 
proliferation of new meson states~\cite{Jaf05}. They would also account for
the possible existence of exotic mesons as could be stable $cc\bar n\bar n$ states,
the topic for discussion since the early 80's~\cite{Ade82}.

Four-quark systems present a richer color structure than standard baryons or mesons.
While the color wave function for standard mesons and baryons leads to a single vector, 
working with four--quark states there are different vectors driving to a singlet color state 
out of colorless or colored quark-antiquark two-body components. Thus, dealing with four--quark states
an important question is whether we are in 
front of a colorless meson-meson molecule or a compact 
state, i.e., a system  with two-body colored components. 
While the first structure would be natural in the naive quark model,
the second one would open a new area on the hadron spectroscopy.

In this manuscript we derive the necessary formalism to evaluate the 
probability of physical channels (singlet--singlet color states) in an
arbitrary four--quark 
wave function. For this purpose one needs to expand any hidden--color vector of the 
four--quark state color basis, i.e., vectors with non--singlet internal color couplings, 
in terms of singlet--singlet color vectors. We will see that  
given a general four--quark state $[q_1q_2\bar q_3\bar q_4]$ the above procedure requires to mix terms 
from two different couplings, $[(q_1\bar q_3)(q_2\bar q_4)]$ and $[(q_1\bar q_4)(q_2\bar q_3)]$. 
If $(q_1,q_2)$ and $(\bar q_3,\bar q_4)$ are identical quarks and antiquarks then, 
a general four-quark wave function can be expanded in terms of color singlet-singlet 
nonorthogonal vectors and therefore the determination of the probability of physical 
channels becomes cumbersome. A particular case has been discussed in the 
literature trying to understand the light scalar mesons 
as $K\bar K$ molecules~\cite{Wei90}. This problem has also been 
found in other fields, as for example in molecular systems~\cite{Man90}.

The manuscript is organized as follows. In Sec.~\ref{tech} the formalism to expand the four--quark 
wave function in terms of singlet--singlet color vectors is derived. Without loss of generality,
the evaluation of the probabilities is exemplified with the 
$QQ\bar n\bar n$ system. In Sec.~\ref{pepe} we discuss some examples 
of bound and unbound states in the charm and bottom sectors. 
Finally, we summarize in Sec.~\ref{summary} our conclusions.

\section{Formalism}
\label{tech}

Given an arbitrary state $|\Psi\rangle$ made up of two quarks,
$q_1$ and $q_2$, and two antiquarks, $\bar q_3$ and  $\bar q_4$, its most 
general wave function will be the direct product of vectors from the 
color, spin, flavor and radial subspaces. We start discussing the color 
substructure.  

\subsection{Color substructure.}
\label{subColor}

There are three different ways of coupling two quarks and two antiquarks 
into a colorless state:
\begin{subequations}
\begin{eqnarray}
\label{eq1a}
[(q_1q_2)(\bar q_3\bar q_4)]&\equiv&\{|\bar 3_{12}3_{34}\rangle,|6_{12}\bar 6_{34}\rangle\}\equiv\{|\bar 33\rangle_c^{12},
|6\bar 6\rangle_c^{12}\}\\
\label{eq1b}
[(q_1\bar q_3)(q_2\bar q_4)]&\equiv&\{|1_{13}1_{24}\rangle,|8_{13} 8_{24}\rangle\}\equiv\{|11\rangle_c,|88\rangle_c\}\\
\label{eq1c}
[(q_1\bar q_4)(q_2\bar q_3)]&\equiv&\{|1_{14}1_{23}\rangle,|8_{14} 8_{23}\rangle\}\equiv\{|1'1'\rangle_c,|8'8'\rangle_c\}\,,
\end{eqnarray}
\label{eq1}
\end{subequations}
\noindent
being the three of them orthonormal basis. Each coupling scheme allows to 
define a color basis where the four--quark problem can be solved. 
The first basis, Eq.~(\ref{eq1a}), 
being the most suitable one to deal with the Pauli principle is made 
entirely of vectors containing hidden--color components. The other two, Eqs.~(\ref{eq1b}) and~(\ref{eq1c}), 
are hybrid basis containing singlet--singlet (physical) and octet--octet (hidden--color) 
vectors.

In order to express a four--quark state in terms only of physical components 
it is necessary to define the antiunitary transformation connecting the basis
(\ref{eq1b}) and (\ref{eq1c})~\cite{Sta96}
\begin{eqnarray}
\ka & = & \cos \alpha \kap  + \sin \alpha \kbp \nonumber \\
\kb & = & \sin \alpha \kap  -  \cos \alpha \kbp  \, ,
\label{anti}
\end{eqnarray}
and the projectors on the different vectors:
\begin{eqnarray}
P & = & \ka \ba \nonumber \\
Q & = & \kb\bb  \, ,
\label{Proj1}
\end{eqnarray}
and
\begin{eqnarray}
\hat P & = & \kap\bap \nonumber \\
\hat Q & = & \kbp\bbp  \, .
\label{Proj2}
\end{eqnarray}

\noindent With these definitions any arbitrary state $|\Psi\rangle$ can be written as
\begin{equation}
|\Psi\rangle  =  P|\Psi\rangle  +  Q|\Psi\rangle =\hat P|\Psi\rangle  +  \hat Q|\Psi\rangle\,.
\label{ii}
\end{equation}
One can extract the singlet--singlet components from the octet--octet one, 
$Q|\Psi\rangle$, by inserting identities $\openone=P+Q=\hat P+\hat Q$ in 
the following iterative manner:
\begin{eqnarray}
|\Psi\rangle  &=&  P|\Psi\rangle  +  \hat P Q|\Psi\rangle  +  \hat Q Q |\Psi\rangle = \nonumber \\
        &=&  P|\Psi\rangle  +  \hat P Q|\Psi\rangle  +  P\hat Q Q |\Psi\rangle  +  Q\hat Q Q |\Psi\rangle = \nonumber \\
        &=&  P|\Psi\rangle  +  \hat P Q|\Psi\rangle  +  P\hat Q Q |\Psi\rangle  +  \hat P Q\hat Q Q |\Psi\rangle 
	 + 	\hat Q Q \hat Q Q |\Psi\rangle = \ldots = \nonumber \\
        &=&  P \left[ \openone  +  \hat Q Q  +  \hat Q Q \hat Q Q  +  \ldots \right] |\Psi\rangle 
        +  \hat P \left[ Q  +  Q \hat Q Q  +  Q \hat Q Q \hat Q Q  +  \ldots \right] |\Psi\rangle \,.
\label{nn}
\label{pp}
\end{eqnarray}
From the definition of the projectors in Eqs.~(\ref{Proj1}) and (\ref{Proj2}) 
and the antiunitary transformation in Eq.~(\ref{anti}) one can see that
\begin{equation}
Q\hat Q Q  =  \kb \pbbp \pbpb \bb  =  \mid \pbbp \mid^2  Q  =  \cos^2 \alpha \, \,Q\,.
\label{qqq}
\end{equation}
Therefore, Eq. (\ref{pp}) can be rewritten as,
\begin{eqnarray}
|\Psi\rangle  &=&  P |\Psi\rangle  +  P\hat Q Q \left[ 1  +  \cos^2 \alpha  +  \cos^4 \alpha  + \ldots \right] |\Psi\rangle \nonumber \\
        &+&  \hat P Q \left[ 1  +  \cos^2 \alpha  +  \cos^4 \alpha  + \ldots \right] |\Psi\rangle\,.
\label{gg}
\end{eqnarray}
Noting that,
\begin{equation}
\sum\limits_{k=0}^{\infty}\cos \alpha^{2k}=\frac{1}{1-\cos^2\alpha} \, ,
\label{serie}
\end{equation}
Eq. (\ref{gg}) becomes,
\begin{equation}
|\Psi\rangle  =  \left[ P  +  P\hat Q Q \frac{1}{1-\cos^2\alpha} \right] |\Psi\rangle  +  
             \left[ \hat P Q \frac{1}{1-\cos^2\alpha} \right] |\Psi\rangle\,.
\label{bj}
\end{equation}
This equation can be simplified by considering 
\begin{equation}
P\hat Q P = \ka \pabp \pbpa \ba = \sin^2 \alpha  \,\, P
\end{equation}
and thus
\begin{eqnarray}
P+P\hat Q Q \frac{1}{1-\cos^2\alpha}&=&P+P\hat Q (1-P) \frac{1}{1-\cos^2\alpha}\\\nonumber
&=&P\hat Q \frac{1}{1-\cos^2\alpha}+P-P\hat Q P \frac{1}{1-\cos^2\alpha}=P\hat Q  \frac{1}{1-\cos^2\alpha}
\label{simpl1}
\end{eqnarray}
Therefore Eq. (\ref{bj}) can be finally written in a compact form as
\begin{equation}
|\Psi\rangle=\frac{1}{1-\cos^2\alpha} P\hat Q|\Psi\rangle+\frac{1}{1-\cos^2\alpha}\hat P Q |\Psi\rangle
={\cal P}_{\ka}^{NH_1}|\Psi\rangle+{\cal P}_{\kap}^{NH_1}|\Psi\rangle\, , 
\label{termino1}
\end{equation}
where ${\cal P}_{\ka}^{NH_1}$ and ${\cal P}_{\kap}^{NH_1}$ are nonhermitian
projection operators on the corresponding singlet-singlet 
subspaces (see Appendix \ref{ap1} for proof of their properties).
This expression demonstrates that any octet--octet color component can be expanded, in general, 
as an infinite sum of singlet--singlet color states~\cite{Har81}.

To obtain hermitian operators one can repeat the same procedure using 
the projectors on $\kap$ and $\kbp$ given in Eq. (\ref{Proj2}),
\begin{eqnarray}
|\Psi\rangle  &=&  \hat Q|\Psi\rangle  +  Q \hat P|\Psi\rangle  +  P \hat P |\Psi\rangle = \nonumber \\
&=&  \hat Q|\Psi\rangle  +  Q \hat P|\Psi\rangle  +  \hat Q P \hat P |\Psi\rangle  +  \hat P P \hat P |\Psi\rangle = \nonumber \\
&=&  \hat Q|\Psi\rangle  +  Q \hat P|\Psi\rangle  +  \hat Q P \hat P |\Psi\rangle  +  Q \hat P P \hat P |\Psi\rangle +
\hat P P \hat P |\Psi\rangle = \ldots =\nonumber \\
&=&  \hat Q \left[ \openone  +  P \hat P  +  P \hat P P \hat P   +  \ldots \right] |\Psi\rangle +
        Q \left[ \hat P  +  \hat P P \hat P  +  \hat P P \hat P P \hat P  +  \ldots \right] |\Psi\rangle  \, ,
\label{pp2}
\label{nn2}
\end{eqnarray}
where
\begin{equation}
\hat P P \hat P  =  \kap \papa \paap \bap  =  \mid \papa \mid^2  \hat P  =  \cos^2 \alpha \,\, \hat P\,,
\end{equation}
what allows to rewrite Eq.~(\ref{pp2})
\begin{eqnarray}
|\Psi\rangle  &=&  \hat Q |\Psi\rangle  +  \hat Q P \hat P  \left[ 1  +  \cos^2 \alpha  +  \cos^4 \alpha  + \ldots \right] |\Psi\rangle \nonumber \\
        &+&  Q \hat P \left[ 1  +  \cos^2 \alpha  +  \cos^4 \alpha  + \ldots \right] |\Psi\rangle \,.
\label{gg2}
\end{eqnarray}
Using Eq. (\ref{serie}) one can finally write
\begin{equation}
|\Psi\rangle  =  \left[ \hat Q  +  \hat Q P \hat P \frac{1}{1-\cos^2\alpha} \right] |\Psi\rangle  +  
             \left[ Q \hat P \frac{1}{1-\cos^2\alpha} \right] |\Psi\rangle\,.
\label{bj2}
\end{equation}
This equation can be simplified noting that
\begin{equation}
\hat Q P \hat Q = \kbp \pbpa \pabp \bbp = \sin^2 \alpha  \,\hat Q\,,
\end{equation}
then
\begin{eqnarray}
\hat Q+\hat Q P \hat P \frac{1}{1-\cos^2\alpha}&=&\hat Q+\hat Q P (1-\hat Q) \frac{1}{1-\cos^2\alpha}\\\nonumber
&=&\hat Q P \frac{1}{1-\cos^2\alpha}+\hat Q-\hat Q P \hat Q \frac{1}{1-\cos^2\alpha}=\hat Q P \frac{1}{1-\cos^2\alpha}  \, .
\label{simpl2}
\end{eqnarray}
arriving to the compact notation
\begin{equation}
|\Psi\rangle=\frac{1}{1-\cos^2\alpha} \hat Q P|\Psi\rangle+\frac{1}{1-\cos^2\alpha}Q \hat P |\Psi\rangle
={\cal P}_{\ka}^{NH_2}|\Psi\rangle+{\cal P}_{\kap}^{NH_2}|\Psi\rangle\,,
\label{termino2}
\end{equation}
where ${\cal P}_{\ka}^{NH_2}$ and ${\cal P}_{\kap}^{NH_2}$ are nonhermitian projection operators 
on the corresponding singlet-singlet subspaces.

Combining Eqs. (\ref{termino1}) and (\ref{termino2}) one can write any arbitrary state in the following form,
\begin{eqnarray}
|\Psi\rangle  & = & \frac{1}{2} \left\{ \left[ P\hat Q \frac{1}{1-\cos^2\alpha} \right] |\Psi\rangle  +  
             \left[ \hat P Q \frac{1}{1-\cos^2\alpha} \right] |\Psi\rangle \right\} \nonumber \\
        & + & \frac{1}{2} \left\{ \left[ \hat Q P \frac{1}{1-\cos^2\alpha} \right] |\Psi\rangle  +  
             \left[ Q \hat P \frac{1}{1-\cos^2\alpha} \right] |\Psi\rangle \right\} \, ,
\end{eqnarray}
or equivalently
\begin{eqnarray}
|\Psi\rangle  & = & \frac{1}{2} \left( P\hat Q + \hat Q P \right) \frac{1}{1-\cos^2\alpha}  |\Psi\rangle \nonumber \\ 
        & + & \frac{1}{2} \left( \hat P Q + Q \hat P \right) \frac{1}{1-\cos^2\alpha} |\Psi\rangle\,.
\end{eqnarray}
Thus, one arrives to two hermitian operators that are well--defined 
projectors on the two physical singlet--singlet color states
\begin{eqnarray}
{\cal P}_{\ka} & =&  \left( P\hat Q + \hat Q P \right) \frac{1}{2(1-\cos^2\alpha)} 
\nonumber \\
{\cal P}_{\kap} & =&  \left( \hat P Q + Q \hat P \right) \frac{1}{2(1-\cos^2\alpha)} 
\label{tt}
\end{eqnarray}
and finally,
\begin{equation}
|\Psi\rangle  =  {\cal P}_{\ka} |\Psi\rangle  +  {\cal P}_{\kap} |\Psi\rangle \,.
\label{111p1p}
\end{equation}
Thus, given an arbitrary state $|\Psi\rangle$ its projection on a particular subspace $E$ is given
by $|\Psi\rangle|_E=P_E|\Psi\rangle$. Thus, the probability of finding such an state on this subspace is
\begin{equation}
_E|\langle\Psi|\Psi\rangle|_E=\langle\Psi|P_E^{\dagger}P_E|\Psi\rangle=\langle\Psi|P_E^2|\Psi\rangle=\langle\Psi|P_E|\Psi\rangle\,.
\end{equation}
Therefore, once the projection operators have been constructed, Eq.~(\ref{tt}), the probabilities for finding 
singlet--singlet components are given by,
\begin{eqnarray}
P^{\mid\Psi\rangle}({[11]})&=&\left\langle\Psi\mid{\cal P}_{\ka}\mid\Psi\right\rangle\nonumber\\
P^{\mid\Psi\rangle}({[1'1']})&=&\left\langle\Psi\mid{\cal P}_{\kap}\mid\Psi\right\rangle\, .
\end{eqnarray}
Using Eq.~(\ref{tt}) one arrives to
\begin{eqnarray}
P^{\mid\Psi\rangle}({[11]})&=&\frac{1}{2(1-\cos^2\alpha)} 
\left[ \left\langle\Psi\mid P\hat Q \mid\Psi\right\rangle +
\left\langle\Psi\mid \hat Q P \mid\Psi\right\rangle\right] \nonumber \\
P^{\mid\Psi\rangle}({[1'1']})&=&\frac{1}{2(1-\cos^2\alpha)} 
\left[ \left\langle\Psi\mid \hat P Q \mid\Psi\right\rangle +
\left\langle\Psi\mid Q \hat P \mid\Psi\right\rangle\right] 
\end{eqnarray}
where it can be easily checked that $P^{\mid\Psi\rangle}({[11]})+P^{\mid\Psi\rangle}({[1'1']})=1$.

\subsection{Spin substructure}
\label{sppin}

For a four--quark state one has three different total spins: 0, 1 and 2. The $S_T=2$ case is trivial, because the
basis is one-dimensional. Let us discuss the other two possibilities. For $S_T=0$ the spin basis, in
analogy with Eqs.~(1), are given by:
\begin{subequations}
\begin{eqnarray}
\label{spinbase-a}
[(s_1s_2)_{S_{12}}(s_3s_4)_{S_{34}}]_{0}&\equiv\{|00\rangle^{12}_s,|11\rangle^{12}_s\}\\
\label{spinbase-b}
[(s_1s_3)_{S_{13}}(s_2s_4)_{S_{24}}]_{0}&\equiv\{|00\rangle_s,|11\rangle_s\}\\
\label{spinbase-c}
[(s_1s_4)_{S_{14}}(s_2s_3)_{S_{23}}]_{0}&\equiv\{|0'0'\rangle_s,|1'1'\rangle_s\}
\end{eqnarray}
\label{spinbase}
\end{subequations}
and the corresponding spin projectors
\begin{eqnarray}
P_s&\equiv&|00\rangle_s\,_s\langle00|\\\nonumber
Q_s&\equiv&|11\rangle_s\,_s\langle11|\\\nonumber
\hat P_s&\equiv&|0'0'\rangle_s\,_s\langle0'0'|\\\nonumber
\hat Q_s&\equiv&|1'1'\rangle_s\,_s\langle1'1'|\,.
\end{eqnarray}

It is important to note that the projectors used in the color space
determine the coupling in the spin space. Thus, introducing the corresponding spin projectors 
in Eq. (\ref{111p1p}) one arrives to
\begin{eqnarray}
|\Psi\rangle  &=&  {\cal P}_{\ka} \left(P_s+Q_s\right)|\Psi\rangle  +  {\cal P}_{\kap} \left(\hat P_s+\hat Q_s\right)|\Psi\rangle =
\\\nonumber 
&=&  {\cal P}_{\ka} P_s|\Psi\rangle+{\cal P}_{\ka} Q_s|\Psi\rangle  +  {\cal P}_{\kap}\hat P_s|\Psi\rangle +{\cal P}_{\kap}\hat Q_s|\Psi\rangle \equiv \\\nonumber
&\equiv &{\cal P}_{\ka,|00\rangle_s}|\Psi\rangle+{\cal P}_{\ka,|11\rangle_s}|\Psi\rangle+{\cal P}_{\kap,|0'0'\rangle_s}|\Psi\rangle+{\cal P}_{\kap,|1'1'\rangle_s}|\Psi\rangle\,,
\end{eqnarray}
where ${\cal P}_{\ka,|00\rangle_s}$ and ${\cal P}_{\kap,|0'0'\rangle_s}$ stand for the projectors on
the physical state made up of two $S=0$ $q\bar q$ mesons, and 
${\cal P}_{\ka,|11\rangle_s}$ and ${\cal P}_{\kap,|1'1'\rangle_s}$ for the projectors on
the physical state made up of two $S=1$ $q\bar q$ mesons.

Following our discussion in Subsection~\ref{subColor} the probabilities are given by
\begin{eqnarray}
P[\ka |00\rangle_s]&=&\langle \Psi|{\cal P}_{\ka,|00\rangle_s}|\Psi\rangle\\\nonumber
P[\ka |11\rangle_s]&=&\langle \Psi|{\cal P}_{\ka,|11\rangle_s}|\Psi\rangle\\\nonumber
P[\kap |0'0'\rangle_s]&=&\langle \Psi|{\cal P}_{\kap,|0'0'\rangle_s}|\Psi\rangle\\\nonumber
P[\kap |1'1'\rangle_s]&=&\langle \Psi|{\cal P}_{\kap,|1'1'\rangle_s}|\Psi\rangle,
\end{eqnarray}
and therefore, the total probabilities of finding a physical state made up of two $S=0$ 
$q\bar q$ mesons will be given by
\begin{equation}
P_{MM}=P[\ka |00\rangle_s]+P[\kap |0'0'\rangle_s]\,
\label{prob1}
\end{equation}
and correspondingly the total probability of a physical state made up of two $S=1$ $q\bar q$ 
states
\begin{equation}
P_{M^*M^*}=P[\ka |11\rangle_s]+P[\kap |1'1'\rangle_s]\,.
\label{prob2}
\end{equation}

In the $S_T=1$ case the spin basis are
\begin{subequations}
\begin{eqnarray}
\label{s1basis-a}
[(s_1s_2)_{S_{12}}(s_3s_4)_{S_{34}}]_1&\equiv\{|01\rangle^{12}_s,|10\rangle^{12}_s,|11\rangle^{12}_s\}\\
\label{s1basis-b}
[(s_1s_3)_{S_{13}}(s_2s_4)_{S_{24}}]_1&\equiv\{|01\rangle_s,|10\rangle_s,|11\rangle_s\}\\
\label{s1basis-c}
[(s_1s_4)_{S_{14}}(s_2s_3)_{S_{23}}]_1&\equiv\{|0'1'\rangle_s,|1'0'\rangle_s,|1'1'\rangle_s\}
\end{eqnarray}
\label{s1basis}
\end{subequations}
and the corresponding projectors,
\begin{eqnarray}
P_s&\equiv&|01\rangle_s\,_s\langle01|\\\nonumber
Q_s&\equiv&|10\rangle_s\,_s\langle10|\\\nonumber
W_s&\equiv&|11\rangle_s\,_s\langle11|\\\nonumber
\hat P_s&\equiv&|0'1'\rangle_s\,_s\langle0'1'|\\\nonumber
\hat Q_s&\equiv&|1'0'\rangle_s\,_s\langle1'0'|\\\nonumber
\hat W_s&\equiv&|1'1'\rangle_s\,_s\langle1'1'|\,.
\end{eqnarray}
Following the same procedure as in the $S_T=0$ case one arrives to
\begin{eqnarray}
|\Psi\rangle  &=&  {\cal P}_{\ka} \left(P_s+Q_s+W_s\right)|\Psi\rangle + {\cal P}_{\kap} \left(\hat P_s+\hat Q_s+\hat W_s\right)|\Psi\rangle = \\ \nonumber
&=&  {\cal P}_{\ka} P_s|\Psi\rangle+{\cal P}_{\ka} Q_s|\Psi\rangle+{\cal P}_{\ka} W_s|\Psi\rangle \\\nonumber
&+&  {\cal P}_{\kap}\hat P_s|\Psi\rangle +{\cal P}_{\kap}\hat Q_s|\Psi\rangle+{\cal P}_{\kap}\hat W_s|\Psi\rangle \equiv \\\nonumber
&\equiv &{\cal P}_{\ka,|01\rangle_s}|\Psi\rangle+{\cal P}_{\ka,|10\rangle_s}|\Psi\rangle+{\cal P}_{\ka,|11\rangle_s}|\Psi\rangle\\\nonumber
&+&{\cal P}_{\kap,|0'1'\rangle_s}|\Psi\rangle+{\cal P}_{\kap,|1'0'\rangle_s}|\Psi\rangle+{\cal P}_{\kap,|1'1'\rangle_s}|\Psi\rangle\,,
\end{eqnarray}
where ${\cal P}_{\ka,|01\rangle_s}$, ${\cal P}_{\kap,|0'1'\rangle_s}$,
${\cal P}_{\ka,|10\rangle_s}$, and ${\cal P}_{\kap,|1'0'\rangle_s}$ stand for 
the projectors on the physical state made up of one $S=0$ and one $S=1$ $q\bar q$ mesons 
and ${\cal P}_{\ka,|11\rangle_s}$ and ${\cal P}_{\ka,|1'1'\rangle_s}$ for the projectors on
the physical state made up of two $S=1$ $q\bar q$ mesons.

Finally, the probabilities can be expressed as 
\begin{eqnarray}
P[\ka |01\rangle_s]&=&\langle \Psi|{\cal P}_{\ka,|01\rangle_s}|\Psi\rangle\\\nonumber
P[\ka |10\rangle_s]&=&\langle \Psi|{\cal P}_{\ka,|10\rangle_s}|\Psi\rangle\\\nonumber
P[\ka |11\rangle_s]&=&\langle \Psi|{\cal P}_{\ka,|11\rangle_s}|\Psi\rangle\\\nonumber
P[\kap |0'1'\rangle_s]&=&\langle \Psi|{\cal P}_{\kap,|0'1'\rangle_s}|\Psi\rangle\\\nonumber
P[\kap |1'0'\rangle_s]&=&\langle \Psi|{\cal P}_{\kap,|1'0'\rangle_s}|\Psi\rangle\\\nonumber
P[\kap |1'1'\rangle_s]&=&\langle \Psi|{\cal P}_{\kap,|1'1'\rangle_s}|\Psi\rangle\,,
\end{eqnarray}
and therefore, the total probability of a physical state made up of one $S=0$ and 
one $S=1$ $q\bar q$ meson will be given by
\begin{equation}
P_{MM^*}=P[\ka |01\rangle_s]+P[\ka |10\rangle_s]+P[\kap |0'1'\rangle_s]+P[\kap |1'0'\rangle_s]\,
\label{prob3}
\end{equation}
and the total probability of a physical state made up of two $S=1$ $q\bar q$ mesons by
\begin{equation}
P_{M^*M^*}=P[\ka |11\rangle_s]+P[\kap |1'1'\rangle_s]\,.
\label{prob4}
\end{equation}

\subsection{Flavor substructure}
\label{flavor_sub}

The previous discussion about the color and spin substructure is general and valid for any four--quark state. 
For the flavor part one find several different cases depending on the number of
light quarks. Although the present formalism can be applied to any four--quark
state, it becomes much simpler whether distinguishable quarks are present. This would be,
for example, the case of  
the $nQ\bar n\bar Q$ system, where the Pauli principle does not apply. 
In this system the basis (\ref{eq1b}) and (\ref{eq1c}) are distinguishable due to the flavor part,
they correspond to $[(i_1i_4)_{I_{14}}(i_2i_3)_{I_{23}}]_{I}\equiv[(q\bar c)_{1/2}(c\bar q)_{1/2}]_{I}$ and
$[(i_1i_3)_{I_{13}}(i_2i_4)_{I_{24}}]_{I}\equiv[(q\bar q)_I(c\bar c)_0]_{I}$, and therefore they are orthogonal. This 
makes that the probabilities can be evaluated in the usual way for orthogonal basis as has been done 
in Ref.~\cite{Vij08}. 

Non-orthogonal basis are necessary for the following cases: $QQ \bar n \bar Q'$,
$QQ' \bar n \bar n$, $Qn \bar n \bar n$ and $nn \bar n \bar n$ ($Q$ may be equal
to $Q'$) or their corresponding antiparticles. The isospin basis are:

\begin{itemize}
\item $QQ \bar n \bar Q'$ 
\begin{subequations}
\begin{eqnarray}
\label{favorbase-a}
[(i_1i_2)_{I_{12}}(i_3i_4)_{I_{34}}]_{\frac{1}{2}}&\equiv|0\frac{1}{2}\rangle^{12}_f\\
\label{favorbase-b}
[(i_1i_3)_{I_{13}}(i_2i_4)_{I_{24}}]_{\frac{1}{2}}&\equiv|\frac{1}{2}0\rangle_f\\
\label{favorbase-c}
[(i_1i_4)_{I_{14}}(i_2i_3)_{I_{23}}]_{\frac{1}{2}}&\equiv|0'\frac{1}{2}'\rangle_f\,.
\label{favorbase-d}
\end{eqnarray}
\end{subequations}

\item $QQ' \bar n \bar n$ 
\begin{subequations}
\begin{eqnarray}
\label{flavorbase-a}
[(i_1i_2)_{I_{12}}(i_3i_4)_{I_{34}}]_{I}&\equiv|0I\rangle^{12}_f\\
\label{flavorbase-b}
[(i_1i_3)_{I_{13}}(i_2i_4)_{I_{24}}]_{I}&\equiv|\frac{1}{2}\frac{1}{2}\rangle_f\\
\label{flavorbase-c}
[(i_1i_4)_{I_{14}}(i_2i_3)_{I_{23}}]_{I}&\equiv|\frac{1}{2}'\frac{1}{2}'\rangle_f\,.
\end{eqnarray}
\label{flavorbase}
\end{subequations}

\item $Qn \bar n \bar n$ 
\begin{itemize}
\item $I=1/2$
\begin{subequations}
\begin{eqnarray}
\label{fvorbase-a}
[(i_1i_2)_{I_{12}}(i_3i_4)_{I_{34}}]_{\frac{1}{2}}&\equiv\{ |\frac{1}{2}0\rangle^{12}_f,|\frac{1}{2}1\rangle^{12}_f\}\\
\label{fvorbase-b}
[(i_1i_3)_{I_{13}}(i_2i_4)_{I_{24}}]_{\frac{1}{2}}&\equiv\{ |\frac{1}{2}0\rangle_f,|\frac{1}{2}1\rangle_f\}\\
\label{fvorbase-c}
[(i_1i_4)_{I_{14}}(i_2i_3)_{I_{23}}]_{\frac{1}{2}}&\equiv\{ |\frac{1}{2}'0'\rangle_f,|\frac{1}{2}'1'\rangle_f\}\,.
\label{fvorbase-d}
\end{eqnarray}
\end{subequations}
\item $I=3/2$
\begin{subequations}
\begin{eqnarray}
\label{forbase-a}
[(i_1i_2)_{I_{12}}(i_3i_4)_{I_{34}}]_{\frac{3}{2}}&\equiv|\frac{1}{2}1\rangle^{12}_f\\
\label{forbase-b}
[(i_1i_3)_{I_{13}}(i_2i_4)_{I_{24}}]_{\frac{3}{2}}&\equiv|\frac{1}{2}1\rangle_f\\
\label{forbase-c}
[(i_1i_4)_{I_{14}}(i_2i_3)_{I_{23}}]_{\frac{3}{2}}&\equiv|\frac{1}{2}'1'\rangle_f\,.
\label{forbase-d}
\end{eqnarray}
\end{subequations}
\end{itemize}

\item $nn \bar n \bar n$ 
\begin{itemize}
\item $I=0$
\begin{subequations}
\begin{eqnarray}
\label{pinbase-a}
[(i_1i_2)_{I_{12}}(i_3i_4)_{I_{34}}]_{0}&\equiv\{|00\rangle^{12}_f,|11\rangle^{12}_f\}\\
\label{pinbase-b}
[(i_1i_3)_{I_{13}}(i_2i_4)_{I_{24}}]_{0}&\equiv\{|00\rangle_f,|11\rangle_f\}\\
\label{pinbase-c}
[(i_1i_4)_{I_{14}}(i_2i_3)_{I_{23}}]_{0}&\equiv\{|0'0'\rangle_f,|1'1'\rangle_f\}\, .
\end{eqnarray}
\label{pinbase}
\end{subequations}

\item $I=1$
\begin{subequations}
\begin{eqnarray}
\label{1basis-a}
[(i_1i_2)_{I_{12}}(i_3i_4)_{I_{34}}]_1&\equiv\{|01\rangle^{12}_f,|10\rangle^{12}_f,|11\rangle^{12}_f\}\\
\label{1basis-b}
[(i_1i_3)_{I_{13}}(i_2i_4)_{I_{24}}]_1&\equiv\{|01\rangle_f,|10\rangle_f,|11\rangle_f\}\\
\label{1basis-c}
[(i_1i_4)_{I_{14}}(i_2i_3)_{I_{23}}]_1&\equiv\{|0'1'\rangle_f,|1'0'\rangle_f,|1'1'\rangle_f\}\, .
\end{eqnarray}
\label{1basis}
\end{subequations}

\item $I=2$
\begin{subequations}
\begin{eqnarray}
\label{orbase-a}
[(i_1i_2)_{I_{12}}(i_3i_4)_{I_{34}}]_{2}&\equiv|11\rangle^{12}_f\\
\label{orbase-b}
[(i_1i_3)_{I_{13}}(i_2i_4)_{I_{24}}]_{2}&\equiv|11\rangle_f\\
\label{orbase-c}
[(i_1i_4)_{I_{14}}(i_2i_3)_{I_{23}}]_{2}&\equiv|1'1'\rangle_f\,.
\label{orbase-d}
\end{eqnarray}
\end{subequations}

\end{itemize}
\end{itemize}
For those cases where the basis is one-dimensional the
recoupling among the three different basis introduced in Eq.~(\ref{eq1}) is straightforward. For those
cases where the basis in not one-dimensional one should follow the procedure 
described in Sect.~\ref{sppin}.

\subsection{Radial substructure}

In order to derive the probability of the physical channels one has
finally to analyze the symmetry of the radial wave function. Such analysis will depend
on the particular state chosen. Without loss of generality we will exemplify the procedure
with the particular case of the $QQ\bar n\bar n$ system. 
Any other four--quark system discussed in Sect.~\ref{flavor_sub} could be analyzed
in the same manner. Let us start with the $S_T=0$ state, whose most general wave function reads
\begin{eqnarray}
\label{funcs0}
|\Psi\rangle&=&|R_1\rangle|\bar 33\rangle^{12}_c|00\rangle^{12}_s|0I\rangle^{12}_f+
|R_2\rangle|\bar33\rangle^{12}_c|11\rangle^{12}_s|0I\rangle^{12}_f\\\nonumber&+&
|R_3\rangle|6\bar6\rangle^{12}_c|00\rangle^{12}_s|0I\rangle^{12}_f+
|R_4\rangle|6\bar6\rangle^{12}_c|11\rangle^{12}_s|0I\rangle^{12}_f\, ,
\end{eqnarray}
where $|R_1\rangle,|R_2\rangle,|R_3\rangle$, and $|R_4\rangle$ are radial wave 
functions that due to symmetry properties satisfy 
$\langle R_1|R_1\rangle+\langle R_2|R_2\rangle+\langle R_3|R_3\rangle+\langle R_4|R_4\rangle=1$,
$\langle R_1|R_2\rangle=\langle R_1|R_3\rangle=\langle R_2|R_4\rangle=\langle R_3|R_4\rangle=0$,
$\langle R_1|R_4\rangle\neq0$, and $\langle R_2|R_3\rangle\neq0$. Applying Eqs. 
(\ref{prob1}) and (\ref{prob2}) one obtains
\begin{eqnarray}
P_{MM}&=&P[|11\rangle_c|00\rangle_s]+P[|1'1'\rangle_c|0'0'\rangle_s]\\\nonumber
&=&\frac{1}{4}\left(1+2\langle R_2|R_2\rangle+2\langle R_4|R_4\rangle\right)+\frac{3\sqrt{6}}{8}\left(\langle R_1|R_4\rangle+\langle R_2|R_3\rangle\right)\\\nonumber
P_{M^*M^*}&=&P[|11\rangle_c|11\rangle_s]+P[|1'1'\rangle_c|1'1'\rangle_s]\\\nonumber
&=&\frac{1}{4}\left(1+2\langle R_1|R_1\rangle+2\langle R_3|R_3\rangle\right)-\frac{3\sqrt{6}}{8}\left(\langle R_1|R_4\rangle+\langle R_2|R_3\rangle\right)\,.
\label{pr4cha}
\end{eqnarray}
Finally, the $QQ\bar n\bar n$ $S_T=1$ most general wave function reads
\begin{eqnarray}
\label{funcs1}
|\Psi\rangle&=&|R_1\rangle|\bar 33\rangle^{12}_c|01\rangle^{12}_s|0I\rangle^{12}_f+
|R_2\rangle|\bar33\rangle^{12}_c|10\rangle^{12}_s|0I\rangle^{12}_f+
|R_3\rangle|\bar33\rangle^{12}_c|11\rangle^{12}_s|0I\rangle^{12}_f\\\nonumber
&+&|R_4\rangle|6\bar6\rangle^{12}_c|01\rangle^{12}_s|0I\rangle^{12}_f+
|R_5\rangle|6\bar6\rangle^{12}_c|10\rangle^{12}_s|0I\rangle^{12}_f+
|R_6\rangle|6\bar6\rangle^{12}_c|11\rangle^{12}_s|0I\rangle^{12}_f\, ,
\end{eqnarray}
where $|R_i\rangle$ are radial wave functions that due to symmetry properties 
satisfy $\sum_{i=1}^6\langle R_i|R_i\rangle=1$ and
all the cross products are zero except for $\langle R_1|R_5\rangle$ and $\langle R_2|R_4\rangle$.
Applying Eqs. (\ref{prob3}) and (\ref{prob4}) one gets
\begin{eqnarray}
P_{MM^*}&=&\frac{1}{2}\left(1+\langle R_3|R_3\rangle+\langle R_6|R_6\rangle-\frac{3\sqrt{2}}{2}\left(\langle R_1|R_5\rangle+\langle R_2|R_4\rangle\right)\right)\\\nonumber
P_{M^*M^*}&=&\frac{1}{2}\left(1-\langle R_3|R_3\rangle-\langle R_6|R_6\rangle+\frac{3\sqrt{2}}{2}\left(\langle R_1|R_5\rangle+\langle R_2|R_4\rangle\right)\right)\,.
\label{pr6cha}
\end{eqnarray}

\section{Some illustrative examples}
\label{pepe}

In the previous section we have derived the analytical expansion of an arbitrary
four--quark state wave function in terms of a non-orthogonal basis containing 
only physical channels. The calculation of the probabilities has been exemplified
with the $QQ\bar n \bar n$ system. 

We now apply this formalism to discuss the four-quark nature: unbound, molecular
or compact states, of some illustrative examples. The same discussion could be
done with any other four--quark state
just changing the coupling in isospin space, see Subsect.~\ref{flavor_sub}.
The results we are going to discuss have been obtained solving the
four-body problem by means of a variational method using as trial wave function 
the most general linear combination of gaussians~\cite{Suz98}. 
The accuracy of the variational approach has been tested by comparing
with results obtained by means of the
hyperspherical harmonic expansion~\cite{Vij09}.
Both approaches are in good agreement
The interaction between the quarks is taken from the model of Ref.~\cite{Vij05}.
The same interacting potential used to calculate the four-quark energy is
used to calculate the mass of the thresholds, i.e., the meson masses.

The stability of a four--quark state can be analyzed in terms of $\Delta_E$, 
the energy difference between its mass and that of the 
lowest two-meson threshold,
\begin{equation}
\label{delta}
\Delta_E=E_{4q}-E(M_1,M_2)\, ,
\end{equation}
where $E_{4q}$ stands for the four--quark energy and $E(M_1,M_2)$ for the energy of the 
two-meson threshold. Thus, $\Delta_E<0$ indicates all fall-apart decays 
are forbidden, and therefore one has a proper bound state. $\Delta_E\ge 0$ 
will indicate that the four--quark solution corresponds to 
an unbound threshold (two free mesons). Thus, an energy above the threshold 
would simply mean that the system is unbound within our variational 
approximation, suggesting that the minimum of 
the Hamiltonian is at the two-meson threshold.
\begin{figure}[tb]
\begin{center}
\caption{$H$--type Jacobi vectors. 1,2 stand for heavy quarks and 3,4 for light antiquarks.}
\label{f1}
\epsfig{file=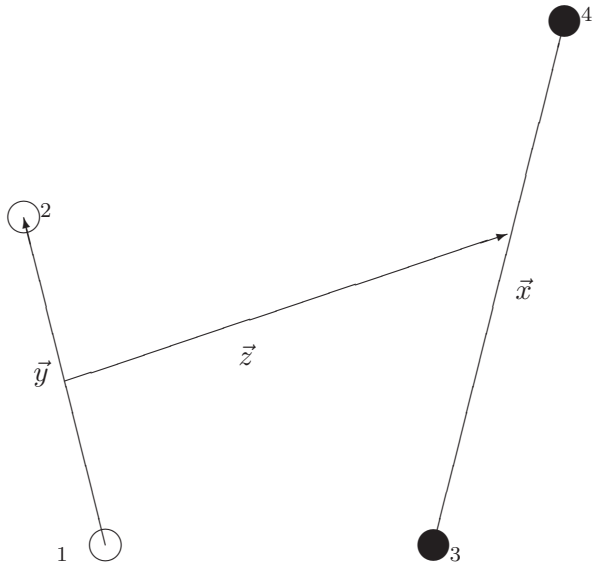}
\end{center}
\end{figure}
Another helpful tool analyzing the structure of a four--quark state is 
the value of the root mean square radii:
$\langle x^2\rangle^{1/2}$, $\langle y^2\rangle^{1/2}$, and $\langle z^2\rangle^{1/2}$. 
They correspond to the Jacobi coordinates given in Fig.\ref{f1}. 
Compact four--quark states can be distinguished from 
two free mesons by means of their root mean square radius 
\begin{equation}
{\rm RMS}_{4q(2q)}= \left(\frac{\sum_{i=1}^{4(2)} m_i \langle (r_i-R)^2\rangle}{\sum_{i=1}^{4(2)} m_i}\right)^{1/2}
\,,
\end{equation}
and in particular, their corresponding ratio,
\begin{equation}
\label{delta-r}
\Delta_R=\frac{{\rm RMS}_{4q}}{{\rm RMS}_{M_1}+{\rm RMS}_{M_2}}\,.
\end{equation}
where ${\rm RMS}_{M_1}+{\rm RMS}_{M_2}$ stands for the sum of the radii of the mesons 
corresponding to the lowest threshold.
\begin{center}
\begin{table}[tb]
\caption{Probabilities for $cc\bar n\bar n$ states with quantum numbers $(S_T,I)$ = (1,1) and (1,0). 
The notation used stands for $P[|\{$Color state$\}\rangle|\{$Spin state$\}\rangle]$ where 
\{Color state\} corresponds to the basis vectors given in Eqs.~(\ref{eq1}) and \{Spin state\} to the ones given in Eqs.~(\ref{s1basis}). Flavor component $|0I\rangle_f$, $|\frac{1}{2}\frac{1}{2}\rangle_f$, and $|\frac{1}{2}'\frac{1}{2}'\rangle_f$ is understood. We list in Appendix~\ref{ap3} a summary of the expressions used.}
\label{tres}
\begin{tabular}{|ccc|ccc|ccc|}
\hline	 	
	 				& (1,1)	& (1,0) && (1,1)	& (1,0) && (1,1)	& (1,0) \\
\hline	 	
$P[|\bar 33\rangle_c^{12}|01\rangle_s]$ & 0.000	& 0.875 
&$P[\ka|01\rangle_s]$ 			& 0.277	& 0.094 
&$P[\kap|0'1'\rangle_s]$ 		& 0.277	& 0.094 \\ 
$P[|\bar 33\rangle_c^{12}|10\rangle_s]$ & 0.000	& 0.006 
&$P[\ka|10\rangle_s]$ 			& 0.277 & 0.094 
&$P[\kap|1'0'\rangle_s]$ 		& 0.277 & 0.094 \\
$P[|\bar 33\rangle_c^{12}|11\rangle_s]$ & 0.333 & 0.000 
&$P[\ka|11\rangle_s]$ 			& 0.002 & 0.186 
&$P[\kap|1'1'\rangle_s]$ 		& 0.002 & 0.186 \\
$P[|6\bar 6\rangle_c^{12}|01\rangle_s]$ & 0.000	& 0.090 
&$P[\kb|01\rangle_s]$ 			& 0.222 & 0.156 
&$P[\kbp|0'1'\rangle_s]$ 		& 0.222 & 0.156 \\
$P[|6\bar 6\rangle_c^{12}|10\rangle_s]$ & 0.000 & 0.029 
&$P[\kb|10\rangle_s]$ 			& 0.222 & 0.156 
&$P[\kbp|1'0'\rangle_s]$ 		& 0.222 & 0.156 \\
$P[|6\bar 6\rangle_c^{12}|11\rangle_s]$	& 0.667	& 0.000 
&$P[\kb|11\rangle_s]$ 			& 0.000 & 0.314 
&$P[\kbp|1'1'\rangle_s]$ 		& 0.000 & 0.314 \\
\hline
$P[|\bar 33\rangle_c^{12}]$ & 0.333 & 0.881 & $P[\ka]$ & 0.556 & 0.374 & $P[\kap]$ & 0.556 & 0.374 \\
$P[|6\bar 6\rangle_c^{12}]$ & 0.667 & 0.119 & $P[\kb]$ & 0.444 & 0.626 & $P[\kbp]$ & 0.444 & 0.626 \\
\hline
\end{tabular}
\end{table}
\end{center}
We show in Table~\ref{tres} the probabilities in color and spin space obtained for 
two $cc\bar n\bar n$ states. The first one, with quantum numbers $(S_T,I) = (1,1)$, 
is unbound, while the second one, $(S_T,I) = (1,0)$, is bound.
We give the probabilities in the three different rearrangements in 
color space defined in Eqs.~(\ref{eq1}). 
Let us note that in the three color rearrangements of Eqs.~(\ref{eq1}), the hidden--color
vectors ($|\bar 33\rangle$, $|6\bar 6\rangle$, $|88\rangle$, and $|8'8'\rangle$) contain
probability of physical channels as we have discussed in Sect.~\ref{tech}.
It is possible to prove from simple group theory arguments that 
for a system composed of two identical quarks ($QQ$) and two identical antiquarks 
($\bar n\bar n$), the octet--octet component probability of the wave 
function either in the (\ref{eq1b}) or (\ref{eq1c}) arrangements
is restricted to the interval $[1/3,2/3]$, see Appendix~\ref{ap2}. 
Do these numerical and analytical results prove the unavoidable presence
of important hidden--color components in all $QQ\bar n\bar n$ states 
regardless of their binding energy? We shall see that the answer is no.
\begin{center}
\begin{table}[tb]
\caption{Four--quark state properties for selected quantum numbers. 
All states have positive parity and total orbital angular momentum $L=0$. 
Energies are given in MeV and distances in fm. The notation 
$M_1M_2\mid_{\ell}$ stands for mesons $M_1$ and $M_2$ with a relative orbital 
angular momentum $\ell$. $P[| \bar 3 3\rangle_c^{12}(| 6\bar 6\rangle_c^{12})]$ stands for the 
probability of the $3\bar 3(\bar 6 6)$ components given in Eq.~(\ref{eq1a}) and $P[\ka(\kb)]$ for the 
$11(88)$ components given in Eq.~(\ref{eq1b}). $P_{MM}$, $P_{MM^*}$, and $P_{M^*M^*}$ have been calculated as 
explained in text.}
\label{t1}
\begin{tabular}{|c|ccccc|}
\hline
$(S_T,I)$ 			& (0,1)	 	 &  (1,1) 	   & (1,0) 	   & (1,0) 	    & (0,0) \\
Flavor 				&$cc\bar n\bar n$&$cc\bar n\bar n$&$cc\bar n\bar n$&$bb\bar n\bar n$&$bb\bar n\bar n$\\
\hline
Energy  			& 3877		 &  3952	   & 3861	   & 10395	    & 10948 \\
Threshold 			& $DD\mid_S$	 &  $DD^*\mid_S$   & $DD^*\mid_S$   & $BB^*\mid_S$   &  $B_1B\mid_P$\\
$\Delta_E$  			& +5		 &  +15		   & $-76$	   & $-$217	    &  $-153$ \\
\hline
$P[| \bar 3 3\rangle_c^{12}]$	& 0.333		 &  0.333	   & 0.881	   & 0.974	    &  0.981 \\
$P[| 6 \bar 6\rangle_c^{12}]$	& 0.667		 &  0.667	   & 0.119	   & 0.026	    &  0.019 \\
\hline
$P[\ka]$			& 0.556		 &  0.556	   & 0.374	   & 0.342	    &  0.340 \\
$P[\kb]$			& 0.444		 &  0.444	   & 0.626	   & 0.658	    &  0.660 \\
\hline
$P_{MM}$ 			& 1.000		 &  $-$		   & $-$	   & $-$	    &  0.254 \\
$P_{MM^*}$ 			& $-$		 &  1.000	   & 0.505	   & 0.531	    &  $-$ \\
$P_{M^*M^*}$ 			& 0.000		 &  0.000	   & 0.495	   & 0.469	    &  0.746 \\
\hline
$\langle x^2\rangle^{1/2}$ 	& 60.988	 &  13.804	   & 0.787	   & 0.684	    &  0.740 \\
$\langle y^2\rangle^{1/2}$ 	& 60.988	 &  13.687	   & 0.590	   & 0.336	    &  0.542 \\
$\langle z^2\rangle^{1/2}$      & 0.433		 &  0.617	   & 0.515	   & 0.503	    &  0.763 \\
$RMS_{4q}$		        & 30.492	 &  6.856	   & 0.363	   & 0.217	    &  0.330 \\
$\Delta_R$			& 69.300	 & 11.640	   &0.799	   & 0.700	    &  0.885 \\
\hline
\end{tabular}
\end{table}
\end{center}
To respond this question we summarize in Table~\ref{t1} the results obtained for 
several different four--quark states, among them those used in Table~\ref{tres},
making use of the formal development of Sect.~\ref{tech}.
As shown in Table~\ref{t1}, independently of their binding energy, all of 
them present a sizable octet-octet 
component when the wave function is expressed in the (\ref{eq1b}) coupling.
Let us first of all concentrate on the 
two unbound states, $\Delta_E > 0$, one with $S_T=0$ and one with $S_T=1$, given
in Table~\ref{t1}. The octet-octet component of basis (1b) can be expanded in terms of
the vectors of basis (1c) as explained in the previous section. Thus,
once expressions (42) and (44) are considered one finds that 
the probabilities are concentrated into a single physical channel, $P_{MM}$ or $P_{MM^*}$.
In other words, the octet-octet component of the basis (1b) or (1c) is a
consequence of having identical quarks and antiquarks. Thus, four-quark 
unbound states are represented by 
two isolated mesons. This conclusion is strengthened when studying 
the root mean square radii,
leading to a picture where the two quarks and the two antiquarks are far 
away, $\langle x^2\rangle^{1/2}\gg 1$ fm and $\langle y^2\rangle^{1/2}\gg 1$ fm, 
while the quark-antiquark pairs are located at a typical distance 
for a meson, $\langle z^2\rangle^{1/2}\le 1$ fm. 

Let us now turn to the bound states shown 
in Table \ref{t1}, $\Delta_E < 0$, one in the charm sector and two in the bottom one. 
Contrary to the results obtained for unbound states, when the octet-octet
component of basis (1b) is expanded in terms of the vectors of basis (1c),
equations (42)
and (44) drive to a picture where the probabilities in all allowed physical channels are 
relevant. It is clear that the bound state must be generated by an interaction
that it is not present in the asymptotic channel, sequestering probability 
from a single singlet--singlet color vector due to the interaction between 
color octets. Such systems are clear examples
of compact four--quark states, in other words, they cannot be expressed in terms of a single physical 
channel. Moreover, as can be seen in Table~\ref{t1}, 
their typical sizes point to compact objects 20\% smaller than a 
standard two--meson system.
\begin{figure}[tb]
\begin{center}
\caption{$P_{MM}$ as a function of $\Delta_E$.}
\label{f2}
\epsfig{file=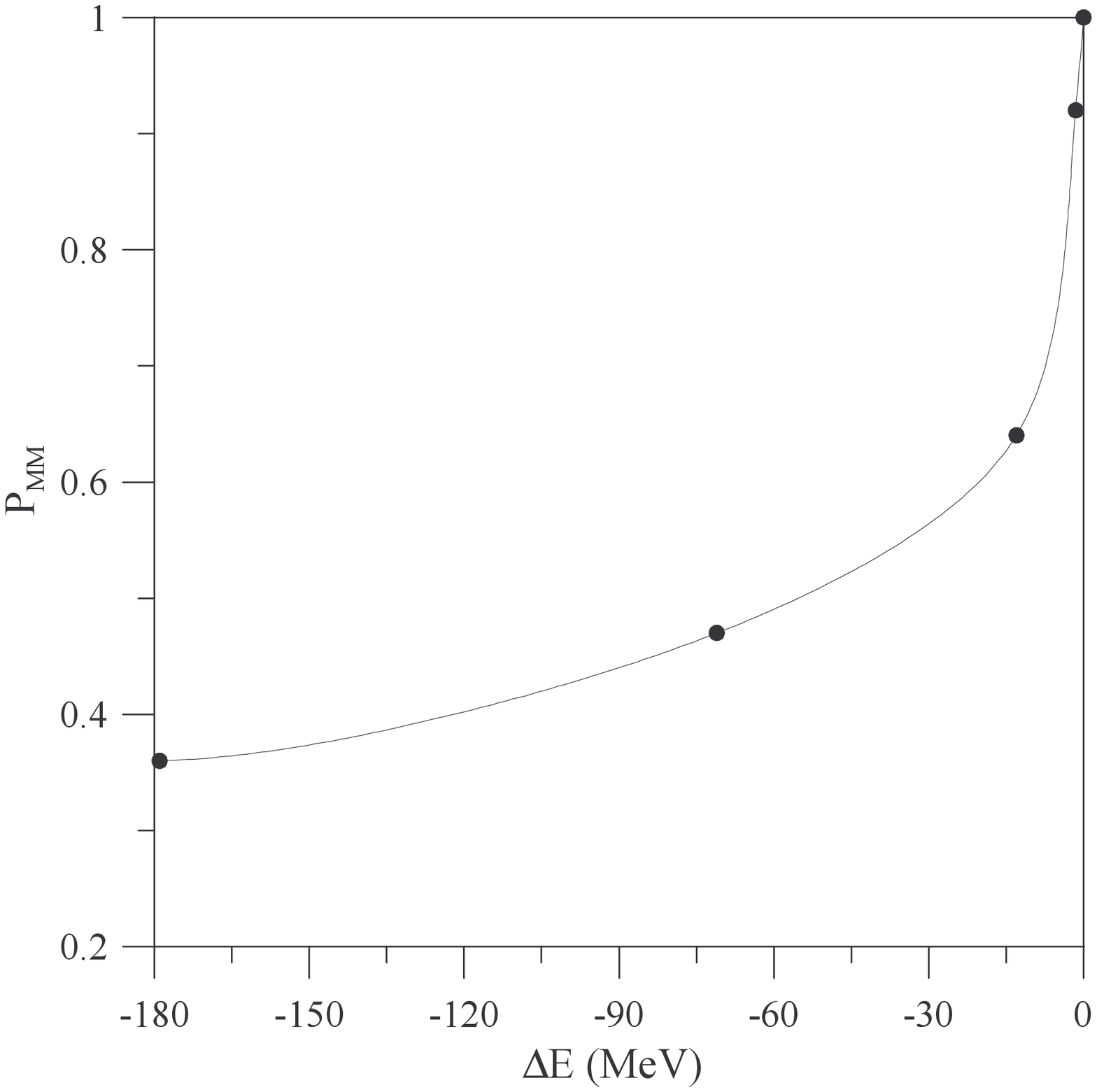, width=8cm}
\end{center}
\end{figure}

\begin{figure}[tb]
\begin{center}
\caption{$\Delta_R$ (a) and  $\langle x^2\rangle^{1/2}$(solid line), $\langle y^2\rangle^{1/2}$ (dashed line), 
and $\langle z^2\rangle^{1/2}$(dashed-dotted line) (b) as a function of $\Delta_E$.}
\label{f4}
\epsfig{file=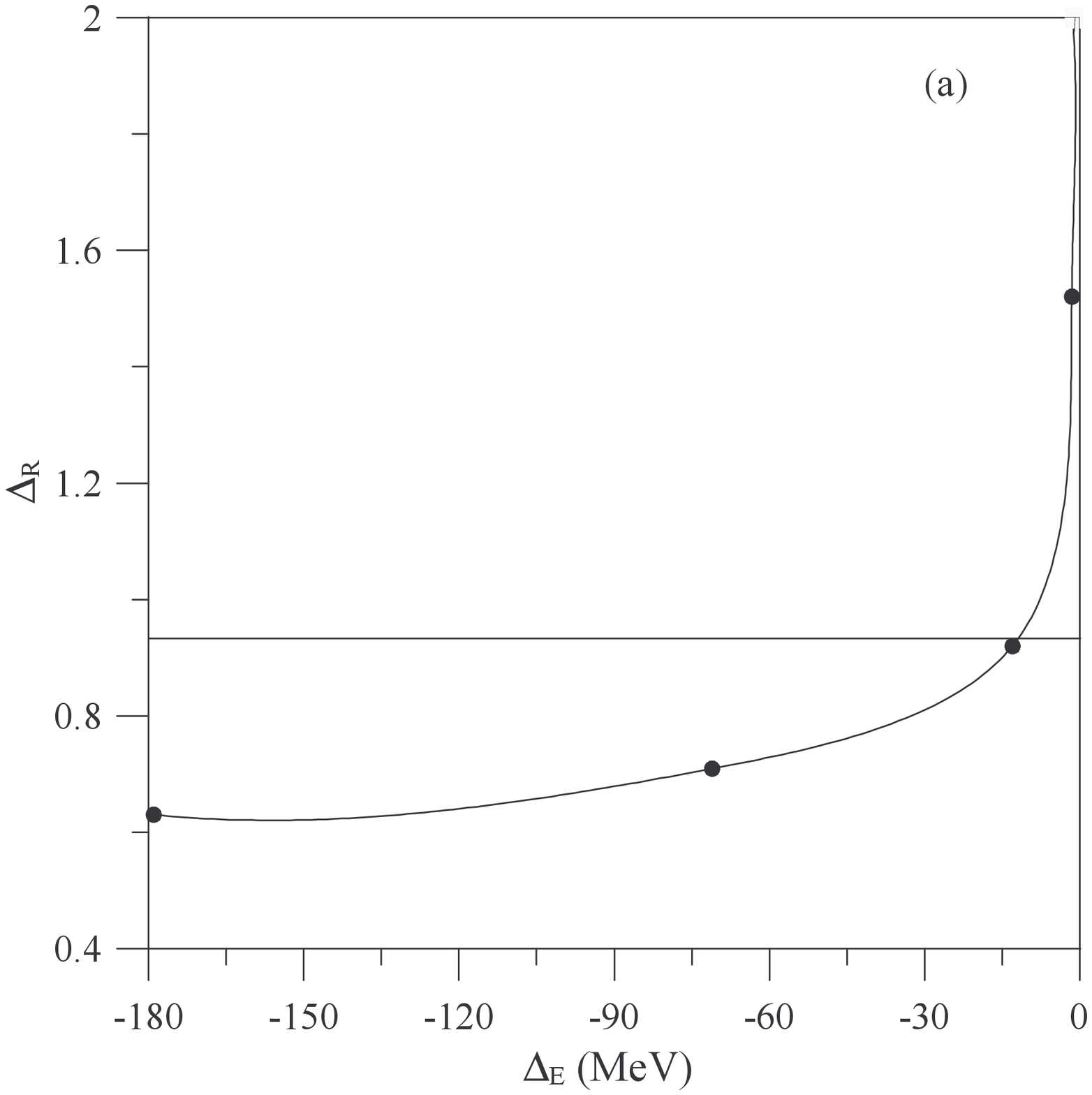, width=7cm}
\hspace*{0.5cm}
\epsfig{file=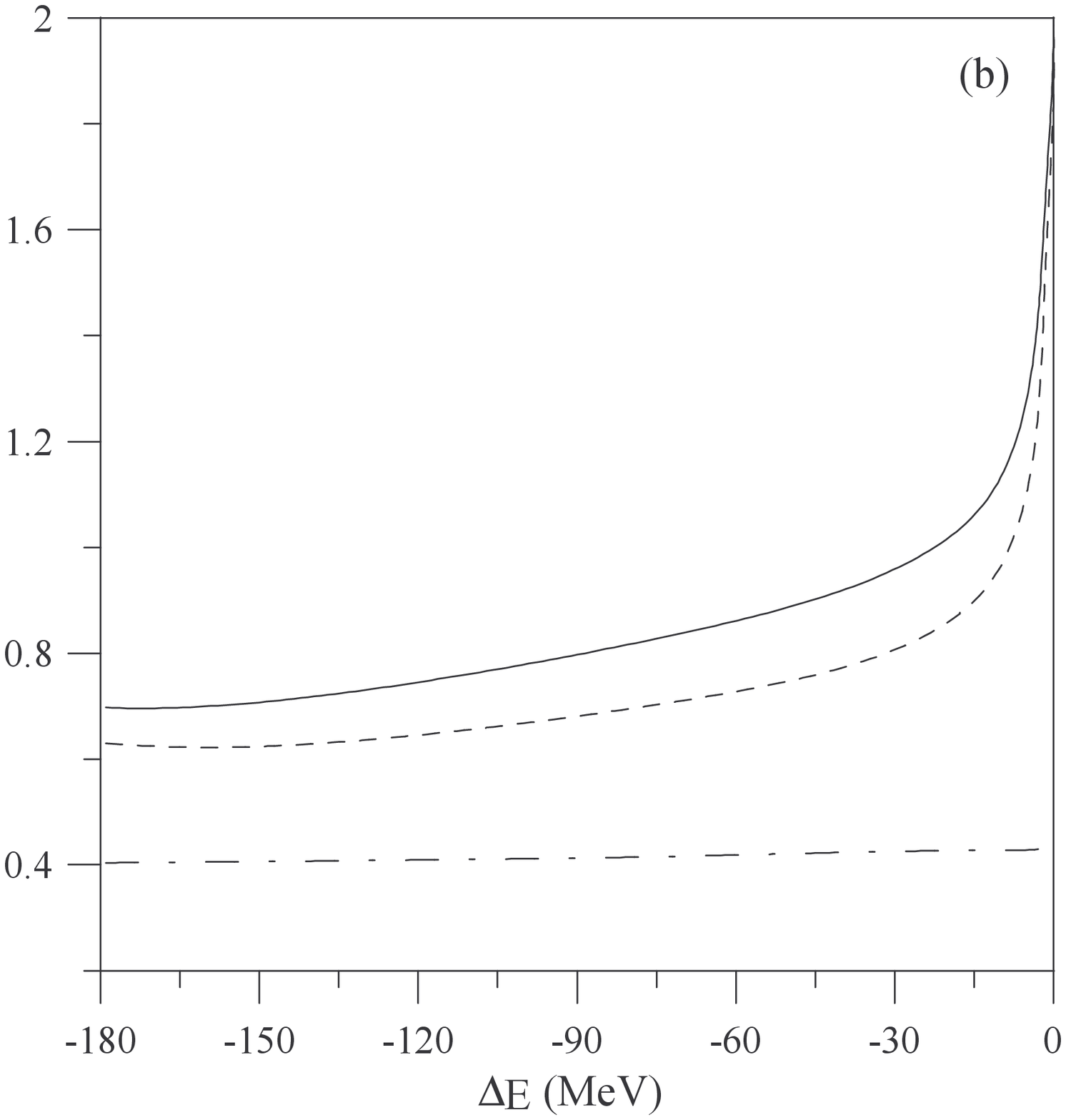, width=6.7cm}
\end{center}
\end{figure}

We have studied the dependence of the probability of a physical channel 
on the binding energy. For this purpose we have considered the simplest
system from the numerical point of view, the 
$(S_T,I)=(0,1)$ $cc\bar n\bar n$ state. Unfortunately, this state
is unbound for any reasonable set of parameters. Therefore, we bind it by multiplying the 
interaction between the light quarks by a fudge factor. 
Such a modification does not affect the two--meson threshold while 
it decreases the mass of the four--quark state. The results are illustrated in 
Fig.~\ref{f2} ($P_{MM}$) and Fig.~\ref{f4} ($\Delta_R$, $\langle x^2\rangle^{1/2},\langle y^2\rangle^{1/2}$, 
and $\langle z^2\rangle^{1/2}$). 
In Fig.~\ref{f2} it is shown how in the $\Delta_E\to0$ limit, 
the four--quark wave function is almost a pure single physical 
channel. Close to this limit one would find what could be defined as 
molecular states.
In Fig.~\ref{f4} we see how the size of the four--quark state increases
when $\Delta_E\to0$. 
It can be observed that when the probability concentrates 
into a single physical channel ($P_{MM}\to 1$) the
size of the system gets larger than the sum of two isolated mesons (Fig.~\ref{f4} left panel).
In Fig.~\ref{f4} (right panel) we identify the subsystems responsible for increasing the size of the four--quark state.
Quark-quark ($\langle x^2\rangle^{1/2}$) and antiquark-antiquark ($\langle y^2\rangle^{1/2}$) 
distances grow rapidly while the quark--antiquark  distance ($\langle z^2\rangle^{1/2}$) 
remains almost constant. This reinforces our previous result, pointing to the appearance 
of two meson like structures whenever the binding energy goes to zero. This illustrative 
example emphasizes the importance of performing a simultaneous analysis both of energy 
and wave function in order to detect bound states in the vicinity of a two-meson threshold.

\begin{figure}[htp]
\caption{Position distribution corresponding to unbound ($S_T=1$, $I=1$, $cc\bar n\bar n$), 
molecular ($S_T=0$, $I=1$, $cc\bar n\bar n$, last point in Fig.\protect\ref{f2}), and 
bound ($S_T=1$, $I=0$, $bb\bar n\bar n$) states.}
\label{colorin}
\epsfig{file=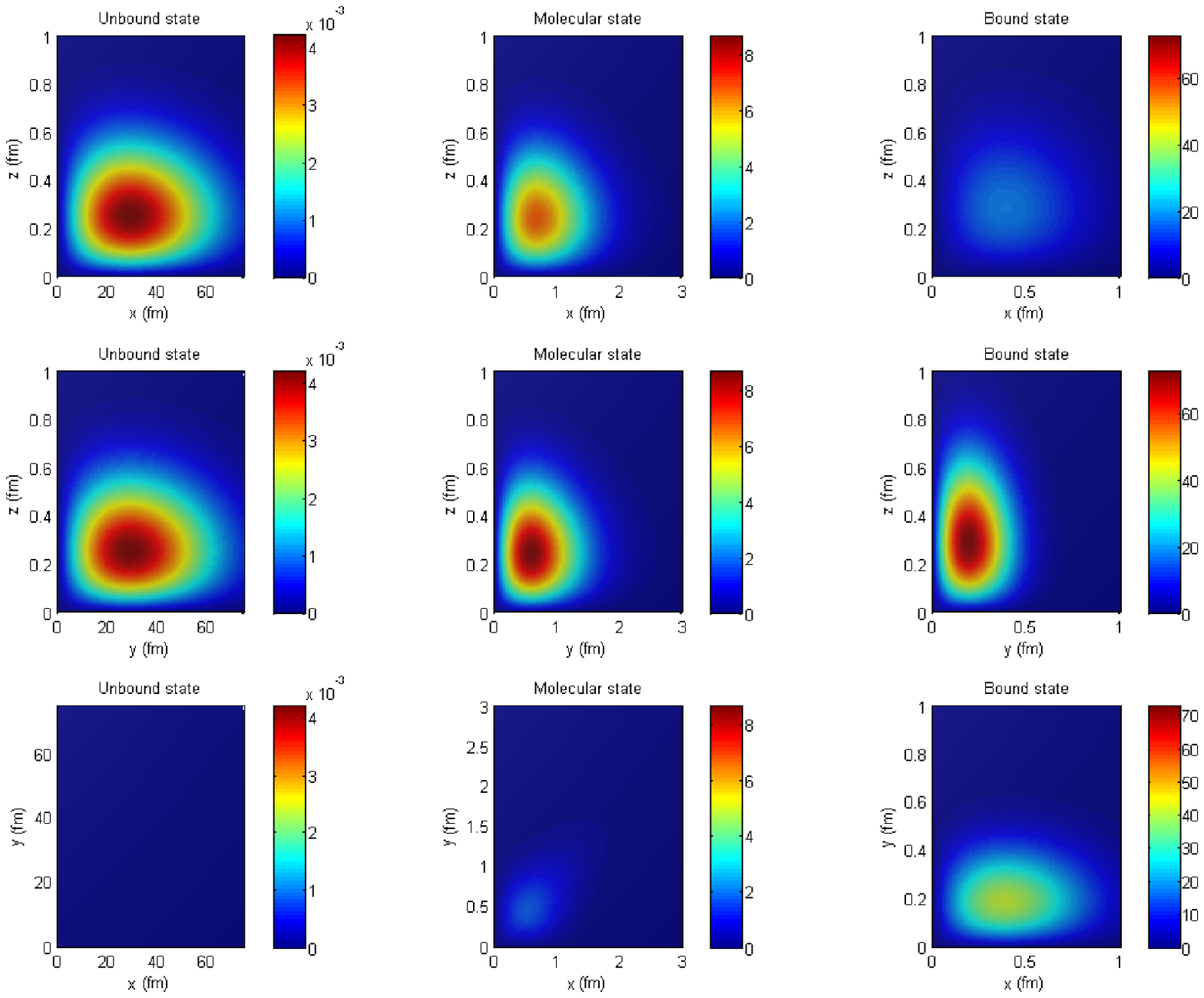, width=18cm}
\end{figure}

To illustrate in more detail the differences observed in the calculated four--quark wave functions we depict in
Fig. \ref{colorin} the position distributions defined as
\begin{equation}
\label{coloreq}
R(r_{\alpha},r_{\beta})=r_{\alpha}r_{\beta}\sum_i\int_V |R_i(\vec r_{\alpha},\vec r_{\beta},\vec r_{\gamma})|^2 d\vec r_{\gamma}\,d\Omega_{r_{\alpha}}\,d\Omega_{r_{\beta}}\,,
\end{equation}
where $R_i(\vec r_{\alpha},\vec r_{\beta},\vec r_{\gamma})$ are the radial wave functions 
introduced in Eqs.~(\ref{funcs0}) and (\ref{funcs1}). We present results for an unbound, a molecular 
and a bound state, showing the position distribution for the different planes 
$(r_{\alpha},r_{\beta})=(z,x),(z,y)$, and $(x,y)$. Clear differences among them 
can be observed. The position distribution for the unbound case spreads in the 
$x$ and $y$ coordinates up to 60 fm, while the bound and molecular systems are 
restricted to the region below 3 fm (molecular) and 1 fm (bound). In the $(x,y)$ 
plane the unbound state is so widely spread that the values for the position 
distribution are three orders of magnitude lower than in the $(z,y)$ and $(z,x)$ cases, 
and therefore they will not appear in the picture unless artificially magnified. 
In the case of the molecular state a long range tail propagating in the $x=y$ region 
can be observed contrary to the constrained values obtained for bound systems.

The conclusions derived are independent of the quark-quark interaction
used. They mainly rely on using the same hamiltonian to describe tensors of
different order, two and four-quark components in the present case. Dealing with a 
complete basis, any four-quark bound deeply bound state has to be compact. Only
slightly bound systems could be considered as molecular. Unbound states correspond
to a two-meson system. A similar situation would
be found in the two baryon system, the deuteron could be considered as a molecular
like state with a small percentage of its wave function on the $\Delta \Delta$ channel,
while the $H-$dibaryon would be a compact six--quark state.
Working with central forces, the only way of getting a bound system is to have 
a strong interaction between the constituents that are far apart in the asymptotic limit,
quarks or antiquarks in the present case. In this case the short-range
interaction will capture part of the probability of a two-meson threshold to form a bound
state. This can be reinterpreted as an infinite sum over physical states. 
This is why
the analysis performed in the present manuscript is so important before concluding
the existence of compact four--quark states beyond simple molecular structures.

If the prescription of using the same hamiltonian to describe all tensors in the Fock space is relaxed,
new scenarios may appear. Among them, the inclusion of many--body forces is particularly relevant.
In Ref.~\cite{Vij07b} the stability of $QQ\bar n\bar n$ and $Q\bar Q n \bar n$ systems 
was analyzed in a simple string model considering only a multiquark confining interaction given 
by the minimum of a flip-flop or a butterfly potential in an attempt to discern whether 
confining interactions not factorizable as two--body potentials would influence the stability 
of four--quark states. The ground state of systems made of two quarks and two antiquarks of 
equal masses was found to be below the dissociation threshold. While for the cryptoexotic 
$Q\bar Q n\bar n$ the binding decreases when increasing the mass ratio $m_Q/m_n$, for the 
flavor exotic $QQ\bar n\bar n$ the effect of mass symmetry breaking is opposite. Others scenarios may emerge
if different many--body forces, like many--body color interactions~\cite{Dmi01} or 't Hooft 
instanton--based three-body interactions~\cite{Hoo76}, are considered.

\section{Conclusions}
\label{summary}

In this work we have developed the necessary formalism to express the wave function of a general four--quark state 
in terms of physical channels, i.e., those constructed by using color singlet states. Once this is done the four--quark 
wave function is expressed in terms of nonorthogonal vectors and hence the traditional way to compute 
probabilities needs to be generalized. We have obtained expressions to evaluate such probabilities for all 
possible nontrivial four--quark states containing two heavy antiquarks and two light quarks. 
We have applied these expressions to illustrative cases, where the difference 
among unbound, compact and molecular four--quark states 
has been made evident. The importance of performing a complete analysis of the system, energy and wave function, in the vicinity of a two-meson threshold has been emphasized.

\acknowledgments
This work has been partially funded by Ministerio de Ciencia y Tecnolog\'{\i}a
under Contract No. FPA2007-65748 and by EU FEDER, and by Junta de Castilla y Le\'{o}n
under Contracts No. SA016A17 and GR12.

\appendix 
\section{Projectors properties}
\label{ap1}

The following properties are proved for one particular set of projectors, 
${\cal P}_{\ka}^{NH_1}$ and ${\cal P}_{\kap}^{NH_1}$.
The same procedure can be followed in all the
remaining cases. By construction, see Eq. (24), they span the complete space, 
${\cal P}_{\ka}^{NH_1}+{\cal P}_{\kap}^{NH_1}=\openone$.
We demonstrate that we have constructed idempotent operators,
\begin{eqnarray}
\left({\cal P}_{\ka}^{NH_1}\right)^2&=&\left(\frac{1}{1-\cos^2\alpha}\right)^2 P\hat QP\hat Q\\ \nonumber
&=&\left(\frac{1}{1-\cos^2\alpha}\right)^2\kap \papb \pbap \papb \bb\\ \nonumber
&=&\left(\frac{1}{1-\cos^2\alpha}\right)^2\kap \papb \bb |\papb|^2 \\ \nonumber
&=&\left(\frac{1}{1-\cos^2\alpha}\right)^2\kap \papb \bb\sin^2\alpha\\ \nonumber
&=&\frac{1}{1-\cos^2\alpha}P\hat Q={\cal P}_{\ka}^{NH_1}\,.
\end{eqnarray}

\section{Minimum and maximum value for the octet--octet component probability.}
\label{ap2}

Without loss of generality we consider the $S_T=0$ case. The Pauli Principle requires 
that the radial wave functions $|R_i\rangle$ in Eq.~(41) 
have well-defined permutation properties under the exchange of quarks and 
that of antiquarks, i.e., symmetric ($S$) or antisymmetric ($A$). 
Hence, $|R_1\rangle$ must be antisymmetric under the exchange of the
identical quarks and also under the exchange of antiquarks what we will
denote by $|R_1(AA)\rangle$. Similarly for the other components:
$|R_2(SS)\rangle$, $|R_3(SS)\rangle$, and $|R_4(AA)\rangle$. 
The transformations from (\ref{eq1a}) to (\ref{eq1b}) and from (\ref{spinbase-a}) 
to (\ref{spinbase-b}) are
\begin{eqnarray}
|\bar 33\rangle_c^{12}&=&\frac{1}{\sqrt{3}}\ka-\sqrt{\frac{2}{3}}\kb\\\nonumber
|6\bar 6\rangle_c^{12}&=&\sqrt{\frac{2}{3}}\ka+\frac{1}{\sqrt{3}}\kb
\end{eqnarray}
and
\begin{eqnarray}
|00\rangle^{12}_s&=&\frac{1}{2}|00\rangle_s+\frac{\sqrt{3}}{2}|11\rangle_s\\\nonumber
|11\rangle^{12}_s&=&\frac{\sqrt{3}}{2}|00\rangle_s-\frac{1}{2}|11\rangle_s
\end{eqnarray}
and therefore Eq.~(41) can be written as
\begin{eqnarray}
|\Psi\rangle&=&
  \ka|00\rangle_s|\frac{1}{2}\frac{1}{2}\rangle_f\frac{1}{2\sqrt{3}}\left\{
|R_1\rangle+\sqrt{3}|R_2\rangle+\sqrt{2}|R_3\rangle+\sqrt{6}|R_4\rangle\right\}+\\\nonumber
&&\ka|11\rangle_s|\frac{1}{2}\frac{1}{2}\rangle_f\frac{1}{2\sqrt{3}}\left\{
\sqrt{3}|R_1\rangle-|R_2\rangle+\sqrt{6}|R_3\rangle-\sqrt{2}|R_4\rangle\right\}+\\\nonumber
&&\kb|00\rangle_s|\frac{1}{2}\frac{1}{2}\rangle_f\frac{1}{2\sqrt{3}}\left\{
-\sqrt{2}|R_1\rangle-\sqrt{6}|R_2\rangle+|R_3\rangle+\sqrt{3}|R_4\rangle\right\}+\\\nonumber
&&\kb|11\rangle_s|\frac{1}{2}\frac{1}{2}\rangle_f\frac{1}{2\sqrt{3}}\left\{
-\sqrt{6}|R_1\rangle+\sqrt{2}|R_2\rangle+\sqrt{3}|R_3\rangle-|R_4\rangle\right\}\,.
\label{funcs0b}
\end{eqnarray}
Considering the symmetry properties of the radial part of the wave function,
the probabilities are calculated as
\begin{eqnarray}
P[\ka|00\rangle_s]&=&\frac{1}{12}\left\{
\langle R_1|R_1\rangle+3\langle R_2|R_2\rangle+2\langle R_3|R_3\rangle+6\langle R_4|R_4\rangle\right.\\\nonumber
&+&\left.2\sqrt{6}\,\,{\rm Re}\,\left(\langle R_1|R_4\rangle+\langle R_2|R_3\rangle\right)\right\}\\\nonumber
P[\ka|11\rangle_s]&=&\frac{1}{12}\left\{
3\langle R_1|R_1\rangle+\langle R_2|R_2\rangle+6\langle R_3|R_3\rangle+2\langle R_4|R_4\rangle\right.\\\nonumber
&-&\left.2\sqrt{6}\,\,{\rm Re}\,\left(\langle R_1|R_4\rangle+\langle R_2|R_3\rangle\right)\right\}\\\nonumber
P[\kb|00\rangle_s]&=&\frac{1}{12}\left\{
2\langle R_1|R_1\rangle+6\langle R_2|R_2\rangle+\langle R_3|R_3\rangle+3\langle R_4|R_4\rangle\right.\\\nonumber
&-&\left.2\sqrt{6}\,\,{\rm Re}\,\left(\langle R_1|R_4\rangle+\langle R_2|R_3\rangle\right)\right\}\\\nonumber
P[\kb|11\rangle_s]&=&\frac{1}{12}\left\{
6\langle R_1|R_1\rangle+2\langle R_2|R_2\rangle+3\langle R_3|R_3\rangle+\langle R_4|R_4\rangle\right.\\\nonumber
&+&\left.2\sqrt{6}\,\,{\rm Re}\,\left(\langle R_1|R_4\rangle+\langle R_2|R_3\rangle\right)\right\}\,.
\label{funcs0c}
\end{eqnarray}
Thus,
\begin{eqnarray}
\label{funcs0d}
P[\ka]&=&P[\ka|00\rangle_s]+P[\ka|11\rangle_s]\\\nonumber
&=&\frac{1}{12}\left\{4\langle R_1|R_1\rangle+4\langle R_2|R_2\rangle+8\langle R_3|R_3\rangle+8\langle R_4|R_4\rangle\right\}\\\nonumber
P[\kb]&=&P[\kb|00\rangle_s]+P[\kb|11\rangle_s]\\\nonumber
&=&\frac{1}{12}\left\{
8\langle R_1|R_1\rangle+8\langle R_2|R_2\rangle+4\langle R_3|R_3\rangle+4\langle R_4|R_4\rangle\right\}\,.
\end{eqnarray}
By construction $P[|\bar 33\rangle_c^{12}]=\langle R_1|R_1\rangle+\langle R_2|R_2\rangle$ and
$P[|6\bar 6\rangle_c^{12}]=\langle R_3|R_3\rangle+\langle R_4|R_4\rangle$ with $P[|\bar 33\rangle_c^{12}]+P[|6\bar 6\rangle_c^{12}]=1$. Therefore Eqs.~(\ref{funcs0d}) can be expressed as
\begin{eqnarray}
\label{funcs0e}
P[\ka]&=&\frac{1}{3}\left\{1+P[|6\bar 6\rangle_c^{12}]\right\}\\\nonumber
P[\kb]&=&\frac{1}{3}\left\{2-P[|6\bar 6\rangle_c^{12}]\right\}\,.
\end{eqnarray}
and since $P[|6\bar 6\rangle_c^{12}]\in[0,1]$ is normalized, a minimum (1/3) and a maximum 
(2/3) value for $P[\ka]$ and $P[\kb]$ do exist.

\section{Probabilities for different choices of basis.}
\label{ap3}

The $S_T=0$ case is given in Eqs.~(\ref{funcs0c}), for the sake of completeness note that
$P[\ka|00\rangle_s]=P[\kap|0'0'\rangle_s]$, $P[\ka|11\rangle_s]=P[\kap|1'1'\rangle_s]$,
$P[\kb|00\rangle_s]=P[\kbp|0'0'\rangle_s]$,$P[\kb|11\rangle_s]=P[\kap|1'1'\rangle_s]$. For $S_T=1$ one has
\begin{eqnarray}
P[\ka|01\rangle_s]&=&P[\kap|0'1'\rangle_s]=\\\nonumber
&&\frac{1}{6}\left(1-\frac{1}{2}\langle R_1|R_1\rangle-\frac{1}{2}\langle R_2|R_2\rangle+\langle R_6|R_6\rangle
-\sqrt{2}[\langle R_1|R_5\rangle+\langle R_2|R_4\rangle]\right)\\\nonumber
P[\ka|10\rangle_s]&=&P[\kap|1'0'\rangle_s]=\\\nonumber
&&\frac{1}{6}\left(1-\frac{1}{2}\langle R_1|R_1\rangle-\frac{1}{2}\langle R_2|R_2\rangle+\langle R_6|R_6\rangle
-\sqrt{2}[\langle R_1|R_5\rangle+\langle R_2|R_4\rangle]\right)\\\nonumber
P[\ka|11\rangle_s]&=&P[\kap|1'1'\rangle_s]=\\\nonumber
&&\frac{1}{6}\left(\langle R_1|R_1\rangle+\langle R_2|R_2\rangle+2\langle R_4|R_4\rangle+2\langle R_5|R_5\rangle
+2\sqrt{2}[\langle R_1|R_5\rangle+\langle R_2|R_4\rangle]\right)\\\nonumber
P[\kb|01\rangle_s]&=&P[\kbp|0'1'\rangle_s]=\\\nonumber
&&\frac{1}{6}\left(1-\frac{1}{2}\langle R_4|R_4\rangle-\frac{1}{2}\langle R_5|R_5\rangle+\langle R_3|R_3\rangle
+\sqrt{2}[\langle R_1|R_5\rangle+\langle R_2|R_4\rangle]\right)\\\nonumber
P[\kb|10\rangle_s]&=&P[\kbp|1'0'\rangle_s]=\\\nonumber
&&\frac{1}{6}\left(1-\frac{1}{2}\langle R_4|R_4\rangle-\frac{1}{2}\langle R_5|R_5\rangle+\langle R_3|R_3\rangle
+\sqrt{2}[\langle R_1|R_5\rangle+\langle R_2|R_4\rangle]\right)\\\nonumber
P[\kb|11\rangle_s]&=&P[\kbp|1'1'\rangle_s]=\\\nonumber
&&\frac{1}{6}\left(2\langle R_1|R_1\rangle+2\langle R_2|R_2\rangle+\langle R_4|R_4\rangle+\langle R_5|R_5\rangle
-2\sqrt{2}[\langle R_1|R_5\rangle+\langle R_2|R_4\rangle]\right)
\end{eqnarray}


\begin{thebibliography}{99}

\bibitem{Bjo85} J.~D.~Bjorken, The November Revolution: A Theorist Reminisces, 
in: A Collection of Summary Talks in High Energy Physics (ed. J.D.~Bjorken),
p. 229 (World Scientific, New York, 2003).

\bibitem{Ros07} J.~L.~Rosner,
		J. Phys. Conf. Ser. {\bf 69}, 012002 (2007).

\bibitem{Jaf05} R.~L.~Jaffe
		Phys. Rep. {\bf 409}, 1 (2005);
		E.~S.~Swanson,
                Phys. Rep. {\bf 429}, 243 (2006);
		E.~Klempt and A.~Zaitsev, 
		Phys. Rep. {\bf 454}, 1 (2007). 

\bibitem{Ade82} J.~P.~Ader, J.~M.~Richard, and P.~Taxil,
                Phys. Rev. D {\bf 25}, 2370 (1982);
                J.~L.~Ballot and J.~M.~Richard,
                Phys. Lett. B {\bf 123}, 449 (1983).

\bibitem{Wei90} J.~Weinstein and N. Isgur,
                Phys. Rev. D {\bf 41}, 2236 (1990).

\bibitem{Man90} R.~S.~Manning and N.~De Leon,
		J. Math. Chem. {\bf 5}, 323 (1990).

\bibitem{Sta96} F.~Stancu, {\it Group Theory in Subnuclear Physics},
                Oxford Stud. Nucl. Phys. {\bf 19} 1 (1996).

\bibitem{Har81} M. Harvey,
	Nucl. Phys. {\bf 352}, 301 (1981).

\bibitem{Vij08}	J.~Vijande, E.~Weissman, N.~Barnea, and A.~Valcarce,
		Phys. Rev. D {\bf 76}, 094022 (2007).

\bibitem{Suz98} Y.~Suzuki and K.~Varga, 
		Lect. Notes Phys. {\bf M54}, 1 (1998).

\bibitem{Vij09} J.~Vijande, A.~Valcarce, and N.~Barnea, 
		Phys. Rev. D, {\bf 79} 074010 (2009).

\bibitem{Vij05} J.~Vijande, F.~Fern\'andez, and A.~Valcarce, 
                J. Phys. G {\bf 31}, 481 (2005).

\bibitem{Vij07b} J.~Vijande, A.~Valcarce, and J.~M.~Richard,
                Phys. Rev. D {\bf 76}, 114013 (2007);
		C. Ay, J.~M. Richard, J. H. Rubinstein,
		Phys. Lett. B{\bf 674}, 227 (2009). 

\bibitem{Dmi01} V. Dmitrasinovic,
		Phys. Lett. B {\bf 499}, 135 (2001);
		Phys.  Rev.  D {\bf 67}, 114007 (2003).

\bibitem{Hoo76} G. 't Hooft, 
		Phys. Rev. D {\bf 14}, 3432 (1976).
\end{thebibliography}
\end{document}